\documentstyle[12pt,epsfig]{article}
\begin{document}
\noindent
\begin{center}
{\Large {\bf $f(R)$ Gravity and Crossing the Phantom Divide
Barrier}}\\ \vspace{2cm}
 ${\bf Yousef~Bisabr}$\footnote{e-mail:~y-bisabr@srttu.edu.}\\
\vspace{.5cm} {\small{Department of Physics, Shahid Rajaee Teacher
Training University,
Lavizan, Tehran 16788, Iran.}}\\
\end{center}
\vspace{1cm}
\begin{abstract}
The $f(R)$ gravity models formulated in Einstein conformal frame
are equivalent to Einstein gravity together with a minimally coupled scalar field. We
shall explore phantom behavior of $f(R)$ models in this frame and
compare the results with those of the usual notion of phantom
scalar field.

\end{abstract}
%~~~~~~~PACS Numbers: 98.80.-k \vspace{3cm}
\section{Introduction}
There are strong observational evidences that the expansion of the
universe is accelerating. These observations are based on type Ia
supernova \cite{super}, cosmic microwave background radiation
\cite{cmbr}, large scale structure surveys \cite{ls} and weak
lensing \cite{wl}. There are two classes of models aim at
explaining this phenomenon:  In the first class, one modifies the
laws of gravity whereby a late-time acceleration is produced. A
family of these modified gravity models is obtained by replacing
the Ricci scalar $R$ in the usual Einstein-Hilbert Lagrangian
density for some function $f(R)$ \cite{carro} \cite{sm}.  In the
second class, one invokes a new matter component usually referred
to as dark energy.  This component is described by an equation of
state parameter $\omega \equiv \frac{p}{\rho}$, namely the ratio
of the homogeneous dark energy pressure over the energy density.
For a cosmic speed up, one should have $\omega < -\frac{1}{3}$
which corresponds to an exotic pressure $p<-\rho/3$. Recent
analysis of the latest and the most reliable dataset (the Gold
dataset \cite{gold}) have indicated that significantly better fits
are obtained by allowing a redshift dependent equation of state
parameter \cite{data}.  In particular, these observations favor
the models that allow the equation of state parameter crossing the
line corresponding to $\omega=-1$, the phantom divide line (PDL),
in the near past. It is therefore important to construct dynamical
models that provide a redshift dependent equation of state
parameter and allow for
crossing the phantom barrier.\\
Most simple models of this kind employ a scalar field coupled
minimally to curvature with negative kinetic energy which referred
to as phantom field \cite{ph} \cite{caldwell}. In contrast to
these models, one may consider models which exhibit phantom
behavior due to curvature corrections to gravitational equations
rather than introducing exotic matter systems.  Recently, there is
a number of attempts to find phantom behavior in $f(R)$ gravity
models.  It is shown that one may realize crossing the PDL in this
framework without recourse to any extra component relating to
matter degrees of freedom with exotic behavior \cite{o} \cite{n}.
Following these attempts, we intend to explore phantom behavior in
some $f(R)$ gravity models which have a viable cosmology, i.e. a
matter-dominated epoch followed by a late-time acceleration. In
contrast to \cite{n}, we shall consider $f(R)$ gravity models in
Einstein conformal frame. It should be noted that mathematical
equivalence of Jordan and Einstein conformal frames does not
generally imply that they are also physically equivalent.  In fact
it is shown that some physical systems can be differently
interpreted in different conformal frames \cite {soko} \cite{no}.
The physical status of the two conformal frames is an open
question which we are not going to address here. Our motivation to
work in Einstein conformal frame is that in this frame, $f(R)$
models consist of Einstein gravity plus an additional dynamical
degree of freedom, the scalar partner of the metric tensor. This
suggests that it is this scalar degree of freedom which drives
late-time acceleration in cosmologically viable $f(R)$ models. We
compare this scalar degree of freedom with the usual notion of
phantom scalar field.  We shall show that behaviors of this scalar
field attributed to $f(R)$ models which allow crossing the PDL are
similar to those of a quintessence field with a negative potential
rather than a phantom with a wrong kinetic term.

~~~~~~~~~~~~~~~~~~~~~~~~~~~~~~~~~~~~~~~~~~~~~~~~~~~~~~~~~~~~~~~~~~~~
\section{Phantom as a Minimally coupled Scalar Field}

The simplest class of models that provides a redshift dependent
equation of state parameter is a scalar field minimally coupled to
gravity whose dynamics is determined by a properly chosen
potential function $V(\varphi)$.  Such models are described by the
Lagrangian density \footnote{We use the unit system $8\pi
G=\hbar=c=1$ and the metric signature $(-,+,+,+)$.}
\begin{equation}
L=\frac{1}{2}\sqrt{-g}(R-\alpha ~g^{\mu\nu}\partial_{\mu}\varphi
\partial_{\nu}\varphi-2V(\varphi))
\label{a1}\end{equation} where $\alpha=+1$ for quintessence and
$\alpha=-1$ for phantom. The distinguished feature of the phantom
field is that its kinetic term enters (\ref{a1}) with opposite
sign in contrast to the quintessence or ordinary matter. The
Einstein field equations which follow (\ref{a1}) are
\begin{equation}
R_{\mu\nu}-\frac{1}{2}g_{\mu\nu}R=T_{\mu\nu}
\label{a2}\end{equation} with
\begin{equation}
T_{\mu\nu}=\alpha~\partial_{\mu}\varphi
\partial_{\nu}\varphi-\frac{1}{2}\alpha~g_{\mu\nu}
\partial_{\gamma}\varphi \partial^{\gamma}\varphi-g_{\mu\nu} V(\varphi)
\label{a3}\end{equation} In a homogeneous and isotropic spacetime,
$\varphi$ is a function of time alone.  In this case, one may
compare (\ref{a3}) with the stress tensor of a perfect fluid with
energy density $\rho_{\varphi}$ and pressure $p_{\varphi}$.  This
leads to the following identifications
\begin{equation}
\rho_{\varphi}=\frac{1}{2}\alpha
\dot{\varphi}^2+V(\varphi)~,~~~~~p_{\varphi}=\frac{1}{2}\alpha
\dot{\varphi}^2-V(\varphi) \label{a4}\end{equation} The equation
of state parameter is then given by
\begin{equation}
\omega_{\varphi}=\frac{\frac{1}{2}\alpha
\dot{\varphi}^2-V(\varphi)}{\frac{1}{2}\alpha
\dot{\varphi}^2+V(\varphi)} \label{a5}\end{equation} In the case
of a quintessence (phantom) field with $V(\varphi)>0$
($V(\varphi)<0$) the equation of state parameter remains in the
range $-1<\omega_{\varphi}<1$.  In the limit of small kinetic term
(slow-roll potentials \cite{slow}), it approaches
$\omega_{\varphi}=-1$ but does not cross this line. The phantom
barrier can be crossed by either a phantom field ($\alpha<0$) with
$V(\varphi)>0$ or a quintessence field ($\alpha>0$) with
$V(\varphi)<0$, when we have $2|V(\varphi)|>\dot{\varphi}^2$.  This
situation corresponds to
\begin{equation}
\rho_{\varphi}>0~~~~~,~~~~~p_{\varphi}<0~~~~~,~~~~~V(\varphi)>0~~~~~~~~~~~~~~~phantom
\label{a51}\end{equation}
\begin{equation}
\rho_{\varphi}<0~~~~~,~~~~~p_{\varphi}>0~~~~~,~~~~~V(\varphi)<0~~~~~~~~~~quintessence\label{a52}\end{equation}
Here it is assumed that the scalar field has a canonical kinetic
term $\pm \frac{1}{2}\dot{\varphi}^2$. It is shown \cite{vik} that
any minimally coupled scalar field with a generalized kinetic term
(k-essence Lagrangian \cite{k}) can not lead to crossing the PDL
through a stable trajectory.  However, there are models that
employ Lagrangians containing multiple fields \cite{multi} or
scalar fields with non-minimall coupling \cite{non} which in
principle can achieve crossing the barrier.\\
There are some remarks to do with respect to  $V(\varphi)<0$
appearing in (\ref{a52}).  In fact, the role of negative
potentials in cosmological dynamics has been recently investigated
by some authors \cite{neg}.  One of the important points about the
cosmological models containing such potentials is that they predict that the
universe may end in a singularity even if it is not closed. For
more clarification, consider a model containing different kinds of
energy densities such as matter, radiation, scalar fields and so
on.  The Friedmann equation in a flat universe is $H^2 \propto
\rho_{t}$ with $\rho_{t}=\Sigma_{i}\rho_{i}$ being the sum of all
energy densities. It is clear that the universe expands forever if
$\rho_{t}>0$. However, if the contribution of some kind of energy
is negative so that $\rho_{i}<0$, then it is possible to have
$H^2=0$ at finite time and the size of the universe starts to
decrease \footnote{For a more detailed discussion see, e.g.,
\cite{mac}. }. We will return to this issue in the context of $f(R)$ gravity models in the
next
section.\\
The possibility of existing a fluid with a surenegative pressure
($\omega<-1$) leads to problems such as vacuum instability and
violation of energy conditions \cite{carroll}. For a perfect fluid
with energy density $\rho$ and pressure $p$, the weak energy
condition requires that $\rho\geq 0$ and $\rho+p \geq 0$. These
state that the energy density is positive and the pressure is not
too large compared to the energy density.  The null energy
condition $\rho+p\geq 0$ is a special case of the latter and
implies that energy density can be negative if there is a
compensating positive pressure.  The strong energy condition as a
hallmark of general relativity states that $\rho+p \geq 0$ and
$\rho+3p\geq 0$.  It implies the null energy condition and
excludes excessively large negative pressures. The null dominant
energy condition is a statement that $\rho\geq |p|$. The physical
motivation of this condition is to prevent vacuum instability or
propagation of energy outside the light cone. Applying to an
equation of state $p=\omega \rho$ with a constant $\omega$, it
means that $\omega \geq -1$. Violation of all these reasonable
constraints by phantom, gives an unusual feature to this principal
energy component of the universe. There are however some remarks
concerning how these unusual features may be circumvented
\cite{carroll} \cite{mc}.

~~~~~~~~~~~~~~~~~~~~~~~~~~~~~~~~~~~~~~~~~~~~~~~~~~~~~~~~~~~~~~~~~~~~~~~~~~~~~~~~
\section{$f(R)$ Gravity}
Let us consider an $f(R)$ gravity model described by the action
\begin{equation}
S=\frac{1}{2} \int d^{4}x \sqrt{-g}~ f(R) + S_{m}(g_{\mu\nu},
\psi)\label{b1}\end{equation} where $g$ is the determinant of
$g_{\mu\nu}$, $f(R)$ is an unknown function of the scalar
curvature $R$ and $S_{m}$ is the matter action depending on the
metric $g_{\mu\nu}$ and some matter field $\psi$. It is well-known
that these models are equivalent to a scalar field minimally
coupled to gravity with an appropriate potential function.  In
fact, we may use a new set of variables
\begin{equation}
\bar{g}_{\mu\nu} =p~ g_{\mu\nu} \label{b2}\end{equation}
\begin{equation} \phi = \frac{1}{2\beta} \ln p
\label{b3}\end{equation}
 where
$p\equiv\frac{df}{dR}=f^{'}(R)$ and $\beta=\sqrt{\frac{1}{6}}$.
This is indeed a conformal transformation which transforms the
above action in the Jordan frame to the Einstein frame \cite{soko}
\cite{maeda} \cite{wands}
\begin{equation}
S=\frac{1}{2} \int d^{4}x \sqrt{-g}~\{ \bar{R}-\bar{g}^{\mu\nu}
\partial_{\mu} \phi~ \partial_{\nu} \phi -2V(\phi)\} + S_{m}(\bar{g}_{\mu\nu}
e^{2\beta \phi}, \psi) \label{b4}\end{equation} In the Einstein
frame, $\phi$ is a minimally coupled scalar field with a
self-interacting potential which is given by
\begin{equation}
V(\phi(R))=\frac{Rf'(R)-f(R)}{2f'^2(R)} \label{b5}\end{equation}
Note that the conformal transformation induces the coupling of the
scalar field $\phi$ with the matter sector. The strength of this
coupling $\beta$, is fixed to be $\sqrt{\frac{1}{6}}$ and is the
same for
all types of matter fields. \\
Variation of the action (\ref{b4}) with respect to
$\bar{g}_{\mu\nu}$, gives the gravitational field equations
\begin{equation}
\bar{G}_{\mu\nu}=T^{\phi}_{\mu\nu}+\bar{T}^{m}_{\mu\nu}
\label{b6}\end{equation}
 where
\begin{equation}
\bar{T}^{m}_{\mu\nu}=\frac{-2}{\sqrt{-g}}\frac{\delta
S_{m}}{\delta \bar{g}^{\mu\nu}}\label{b7}\end{equation}
\begin{equation}
T^{\phi}_{\mu\nu}=\partial_{\mu} \phi~\partial_{\nu} \phi
-\frac{1}{2}\bar{g}_{\mu\nu} \partial_{\gamma}
\phi~\partial^{\gamma} \phi-V(\phi) \bar{g}_{\mu\nu}
\label{b8}\end{equation} Here $\bar{T}^{m}_{\mu\nu}$ and
$T^{\phi}_{\mu\nu}$ are stress tensors of the matter system and
the minimally coupled scalar field $\phi$, respectively. Comparing
(\ref{a3}) and (\ref{b8}) indicates that  $\alpha=1$ and $\phi$
appears as a normal scalar field.  Thus the equation of state
parameter which corresponds to $\phi$ is given by
\begin{equation}
\omega_{\phi} \equiv
\frac{p_{\phi}}{\rho_{\phi}}=\frac{\frac{1}{2}
\dot{\phi}^2-V(\phi)}{\frac{1}{2} \dot{\phi}^2+V(\phi)}
\label{b9}\end{equation} Inspection of (\ref{b9}) reveals that for
$\omega_{\phi}<-1$, we should have $V(\phi)<0$ and
$|V(\phi)|>\frac{1}{2}\dot{\phi}^2$ which corresponds to
(\ref{a52}). In explicit terms, crossing the
PDL in this case requires that $\phi$ appear as a quintessence
(rather than a phantom) field with a negative potential. \\Here
the scalar field $\phi$ has a geometric nature and is related to
the curvature scalar by (\ref{b3}). One may therefore use
(\ref{b3}) and (\ref{b5}) in the expression (\ref{b9}) to obtain
\begin{equation}
\omega_{\phi}=\frac{3\dot{R}^2
f''^2(R)-\frac{1}{2}(Rf'(R)-f(R))}{3\dot{R}^2
f''^2(R)+\frac{1}{2}(Rf'(R)-f(R))} \label{b10}\end{equation} which
is an expression relating $\omega_{\phi}$ to the function $f(R)$.
It is now possible to use (\ref{b10}) and find the functional forms
of $f(R)$ that fulfill $\omega_{\phi}<-1$. In general, to find
such $f(R)$ gravity models one may start with a particular $f(R)$
function in the action (\ref{b1}) and solve the corresponding
field equations for finding the form of $H(z)$.  One can then use
this function in (\ref{b10}) to obtain $\omega_{\phi}(z)$.
However, this approach is not efficient in view of
complexity of the field equations.  An alternative
approach is to start from the best fit parametrization $H(z)$
obtained directly from data and use this $H(z)$ for a particular
$f(R)$ function in (\ref{b10}) to find $\omega_{\phi}(z)$.  We
will follow the latter approach to find $f(R)$ models that
provide crossing the phantom barrier.\\
We begin with the Hubble parameter $H\equiv \frac{\dot{a}}{a}$.
Its derivative with respect to cosmic time $t$ is
\begin{equation}
\dot{H}=\frac{\ddot{a}}{a}-(\frac{\dot{a}}{a})^2
\label{b11}\end{equation} where $a(t)$ is the scale factor of the
Friedman-Robertson-Walker (FRW) metric.  Combining this with the
definition of the deceleration parameter
\begin{equation}
q(t)=-\frac{\ddot{a}}{aH^2} \label{b12}\end{equation} gives
\begin{equation}
\dot{H}=-(q+1)H^2 \label{b13}\end{equation} One may use
$z=\frac{a(t_{0})}{a(t)}-1$ with $z$ being the redshift, and the
relation (\ref{b12}) to write (\ref{b13}) in its integration form

\begin{equation}
H(z)=H_{0}~exp~[\int_{0}^{z} (1+q(u))d\ln(1+u)]
\label{b14}\end{equation} where the subscript ``0" indicates the
present value of a quantity.  Now if a function $q(z)$ is given,
then we can find evolution of the Hubble parameter. Here we use a
two-parametric reconstruction function characterizing $q(z)$
\cite{wang}\cite{q},
\begin{equation}
q(z)=\frac{1}{2}+\frac{q_{1}z+q_{2}}{(1+z)^2}
\label{b15}\end{equation} where fitting this model to the Gold
data set gives $q_{1}=1.47^{+1.89}_{-1.82}$ and $q_{2}=-1.46\pm
0.43$ \cite{q}. Using this in (\ref{b14}) yields
\begin{equation}
H(z)=H_{0}(1+z)^{3/2}exp[\frac{q_{2}}{2}+\frac{q_{1}z^2-q_{2}}{2(z+1)^2}]
\label{b16}\end{equation} In a spatially flat FRW spacetime
$R=6(\dot{H}+2H^2)$ and therefore $\dot{R}=6(\ddot{H}+4\dot{H}H)$.
In terms of the deceleration parameter we have
\begin{equation}
R=6(1-q)H^2 \label{b17}\end{equation}and
\begin{equation}
\dot{R}=6H^3 \{2(q^2-1)-\frac{\dot{q}}{H}\}
\label{b18}\end{equation} which the latter is equivalent to
\begin{equation}
\dot{R}=6H^3 \{2(q^2-1)+(1+z)\frac{dq}{dz}\}
\label{b19}\end{equation} It is now possible to use (\ref{b15})
and (\ref{b16}) for finding $R$ and $\dot{R}$ in terms of the
redshift. Then for a given $f(R)$ function, the relation
(\ref{b10}) determines the evolution of the equation of state
parameter $\omega_{\phi}(z)$.  \\As an illustration we apply this
procedure to some $f(R)$ functions. Let us first consider the
model \cite{cap} \cite{A}
\begin{equation}
f(R)=R+\lambda R^n \label{b20}\end{equation} in which $\lambda$
and $n$ are constant parameters.  In terms of the values
attributed to these parameters, the model
(\ref{b20}) is divided in three cases \cite{A}. Firstly, when
$n>1$ there is a stable matter-dominated era which does not follow
by an asymptotically accelerated regime. In this case, $n = 2$
corresponds to Starobinsky's inflation and the accelerated phase
exists in the asymptotic past rather than in the future. Secondly,
when $0<n<1$ there is a stable matter-dominated era followed by an
accelerated phase only for $\lambda<0$. Finally, in the case that
$n<0$ there is no accelerated and matter-dominated phases for
$\lambda>0$ and $\lambda<0$, respectively.  Thus the model
(\ref{b20}) is cosmologically viable in the regions of the
parameters space which is given by $\lambda<0$ and $0<n<1$.\\ Due
to complexity of the resulting $\omega_{\phi}(z)$  function, we do
not explicitly write it here and only plot it in Fig.1a for some
parameters.  As the figure shows, there is no phantom behavior and
$\omega_{\phi}(z)$ remains near the line of the cosmological
constant $\omega_{\phi}=-1$.  We also plot $\omega_{\phi}$ in
terms of $n$ and $\lambda$ for $z=1$ in Fig.1b. The figure shows
that $\omega_{\phi}$ remains near unity except for a small region
in
which $-1\leq \omega_{\phi}<0$ and therefore the PDL is never crossed.\\
Now we consider the models presented by Starobinsky \cite{star}
\begin{equation}
f(R)=R-\gamma R_{c} \{1-[1+(\frac{R}{R_{c}})^2]^{-m}\}
\label{b21}\end{equation} and Hu-Sawicki \cite{hs}
\begin{equation}
f(R)=R-\gamma
R_{c}\{\frac{(\frac{R}{R_{c}})^m}{1+(\frac{R}{R_{c}})^m}\}
\label{b22}\end{equation} where $\gamma$, $m$ and $R_{c}$ are
positive constants with $R_{c}$ being of the order of the
presently observed effective cosmological constant.  Using the
same procedure, we can obtain evolution of the equation of state
parameter for both models (\ref{b21}) and (\ref{b22}). We plot the
resulting functions in Fig.2. The figures show that while the model
(\ref{b22}) allows crossing the PDL for
the given values of the parameters, in the model (\ref{b21}) the equation of state parameter
remains near $\omega_{\phi}=-1$. To explore the behavior of the
models in a wider range of the parameters, we also plot $\omega_{\phi}$ in the redshift $z=1$ in Fig.3.\\  It is interesting to consider
violation of energy conditions for the model (\ref{b22}) which can
exhibit phantom behavior.  In Fig.4, we plot some expressions
corresponding to null, weak and strong energy conditions.  As it
is indicated in the figures, the model violates weak and strong
energy conditions while it respects null energy condition for a
period of evolution of the universe.  Moreover, Fig.4a indicates
that $\rho_{\phi}<0$ for some parameters in terms of which the PDL is crossed.
This is in accord with (\ref{a52}) and (\ref{b9}) which require that in order for crossing the PDL,
$\phi$ should be a quintessence field
with a negative potential function.
~~~~~~~~~~~~~~~~~~~~~~~~~~~~~~~~~~~~~~~~~~~~~~~~~~~~~~~~~~~~~~~~~~~~~~~~~~

\section{Concluding Remarks}
We have studied phantom behavior for some $f(R)$ gravity models in
which the late-time acceleration of the universe is realized.
Working in Einstein conformal frame, we separate the scalar degree
of freedom which is responsible for the late-time acceleration.
Comparing this scalar field with the phantom field, we have made
our first observation that the former appears as a minimally coupled quintessence
whose dynamics is characterized
by a negative potential.  The impact of such a negative potential
in cosmological dynamics is that it leads to a collapsing universe
or a big crunch \cite{neg}. As a consequence, the $f(R)$ gravity
models in which crossing the phantom barrier is realized predict
that the universe stops expanding and eventually collapses.  This
is in contrast to phantom scalar fields in which the final stage
of the universe has a divergence of the scale factor
at a finite time, or a big rip \cite{ph} \cite{caldwell}. \\
We have used the reconstruction functions $q(z)$ and $H(z)$
fitting to the Gold data set to find evolution of equation of
state parameter $\omega_{\phi}(z)$ for some cosmologically viable
$f(R)$ models.  We obtained the following results :\\\\
1) The model (\ref{b20}) does not provide crossing the PDL.  It
however allows $\omega_{\phi}$ to be negative for a small region
in the parameters space.  For $n=0$, the expression (\ref{b20})
appears as the Einstein gravity plus a cosmological constant. This
state is indicated in Fig.1b when the equation of state parameter
experiences a sharp decrease to $\omega_{\phi}=-1$.    \\\\
2) We also do not observe phantom behavior in the Starobinsky's
model (\ref{b21}). In the region of the parameters space corresponding to $m>0.5$ the
equation of state parameter decreases to $\omega_{\phi}=-1$ and the
model effectively appears as $\Lambda$CDM.   \\\\
3) The same analysis is fulfilled for Hu-Sawicki's model
(\ref{b22}).  This model exhibits phantom crossing in a small
region of the parameters space as it is indicated in Fig.2b
and Fig.3b. Due to crossing the PDL in this case, we also examine energy
conditions. We find that in contrast to weak and strong energy
conditions which are violated, the null energy condition hold in a
period of the evolution. \\ Although the properties of $\phi$
differ from those of the phantom due to the sign of its kinetic
term, violation of energy conditions remains as a consequence of
crossing the PDL in both cases.  However, the scalar field $\phi$ in our case
should not be interpreted as an exotic matter since it has a
geometric nature characterized by (\ref{b3}).  In fact, taking
$\omega_{\phi}<-1$ as a condition in (\ref{b10}) just leads to some
algebraic relations constraining the explicit form of the $f(R)$
function.
%%%%%%%%%%%%%%%%%%%%%%%%%%%%%%%%%%%%%%%%%%%%%%%%%%%%%%%%%%%%%%%%%%%%%%%%%%%%%%

\newpage
\Large{\bf Figures :} \vspace{2cm}
\begin{figure}[ht]
\begin{center}
\includegraphics[width=0.45\linewidth]{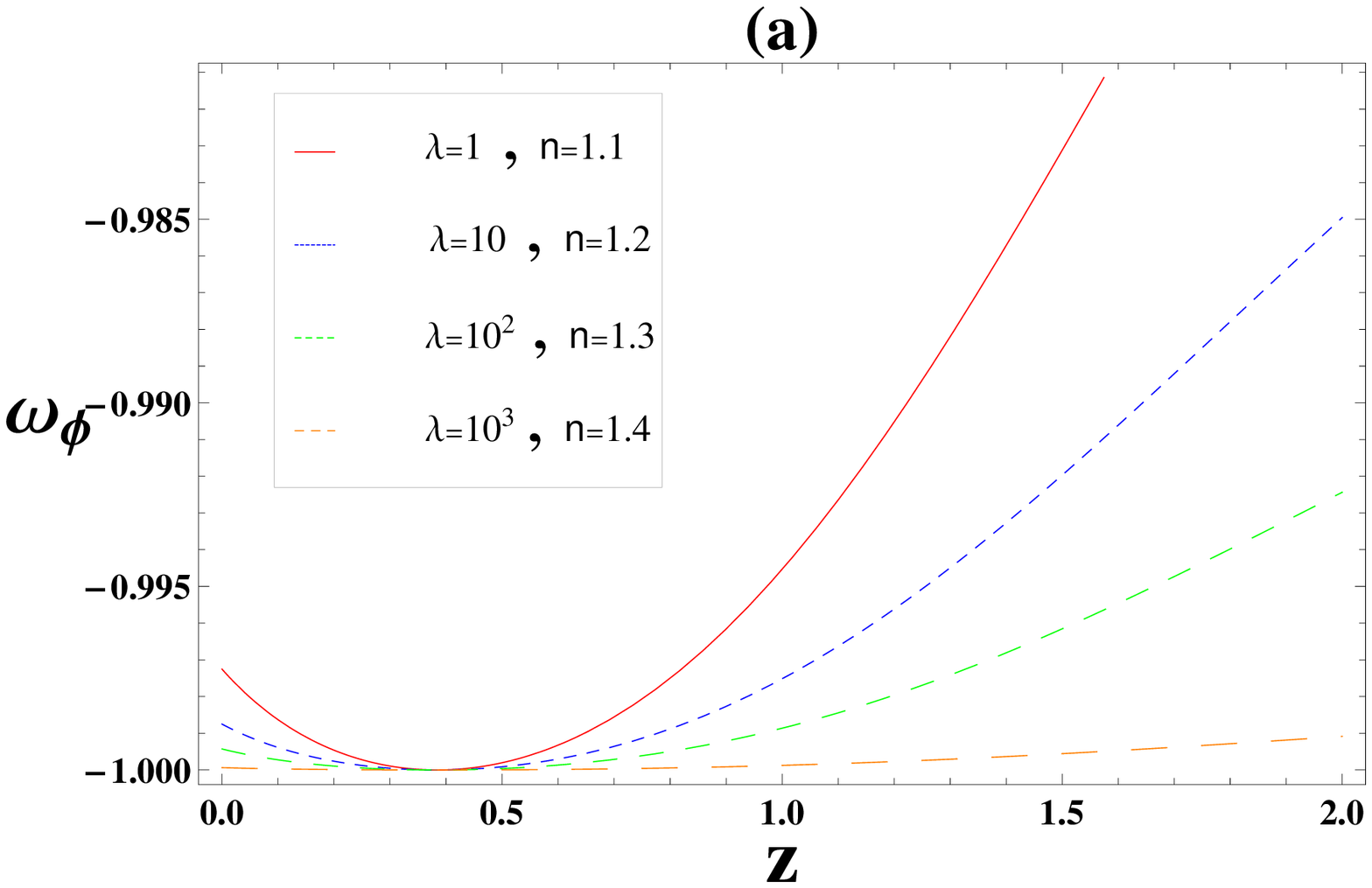}
\includegraphics[width=0.45\linewidth]{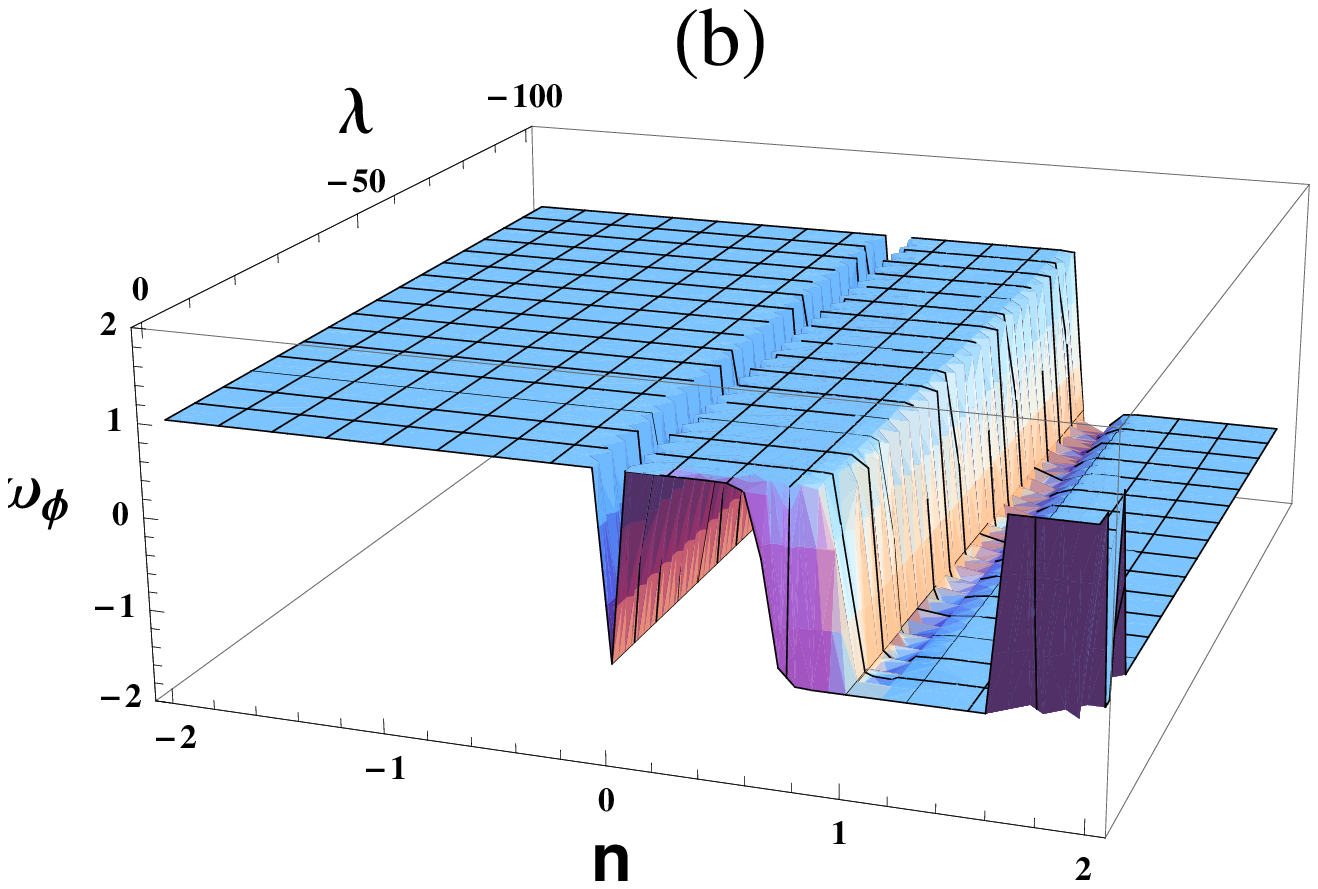}
\caption{a) The plot of $\omega_{\phi}$ in terms of $z$ and for
some values of the parameters $\lambda$ and $n$. There is not any
phantom behavior in these cases.  b) The plot of $\omega_{\phi}$
for the redshift $z=0.25$.  Even though in a small region
$\omega_{\phi}$ takes negative values, it does not however cross
the PDL. }
\end{center}
\end{figure}

\begin{figure}[ht]
\begin{center}
\includegraphics[width=0.49\linewidth]{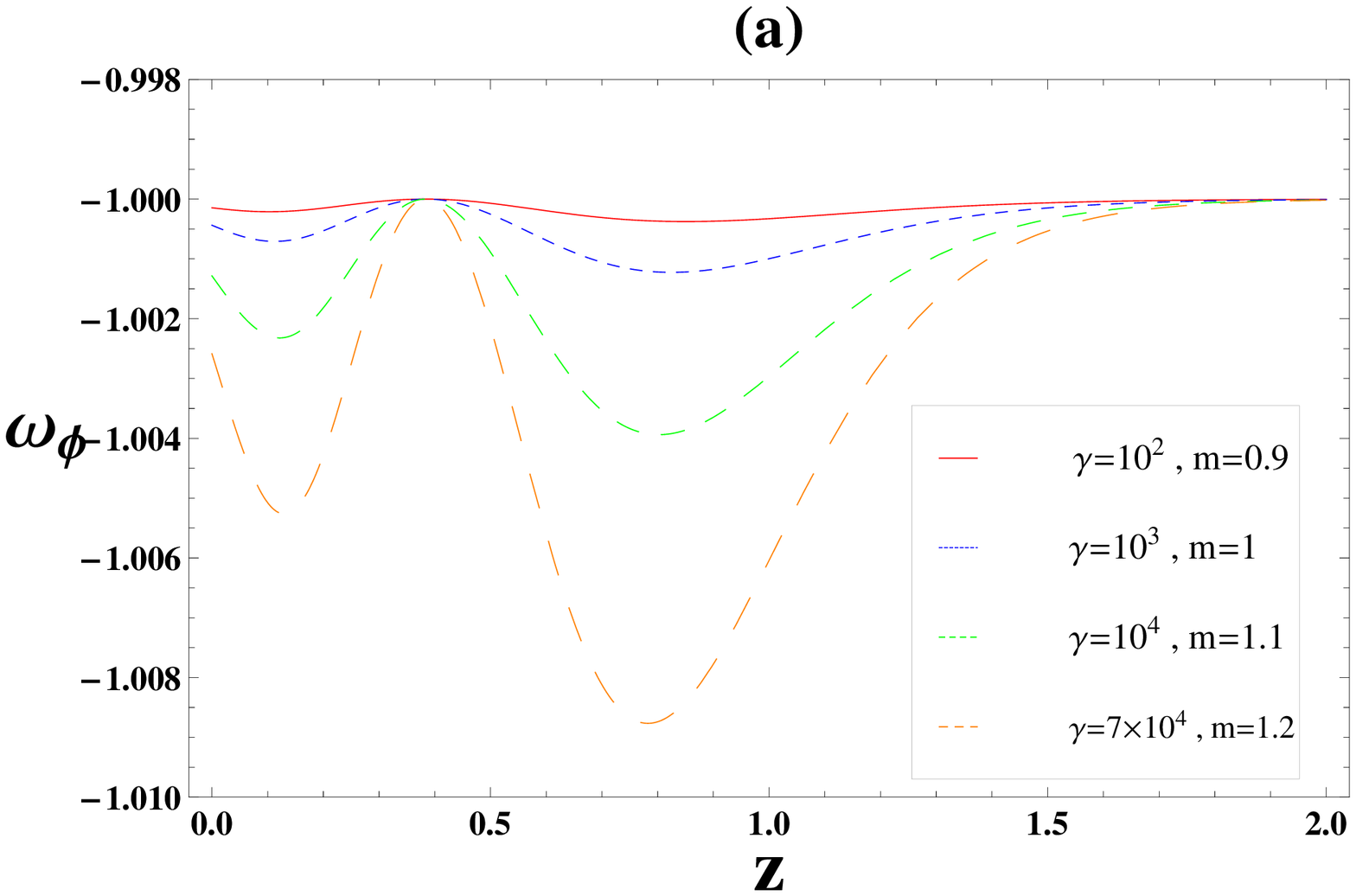}
\includegraphics[width=0.49\linewidth]{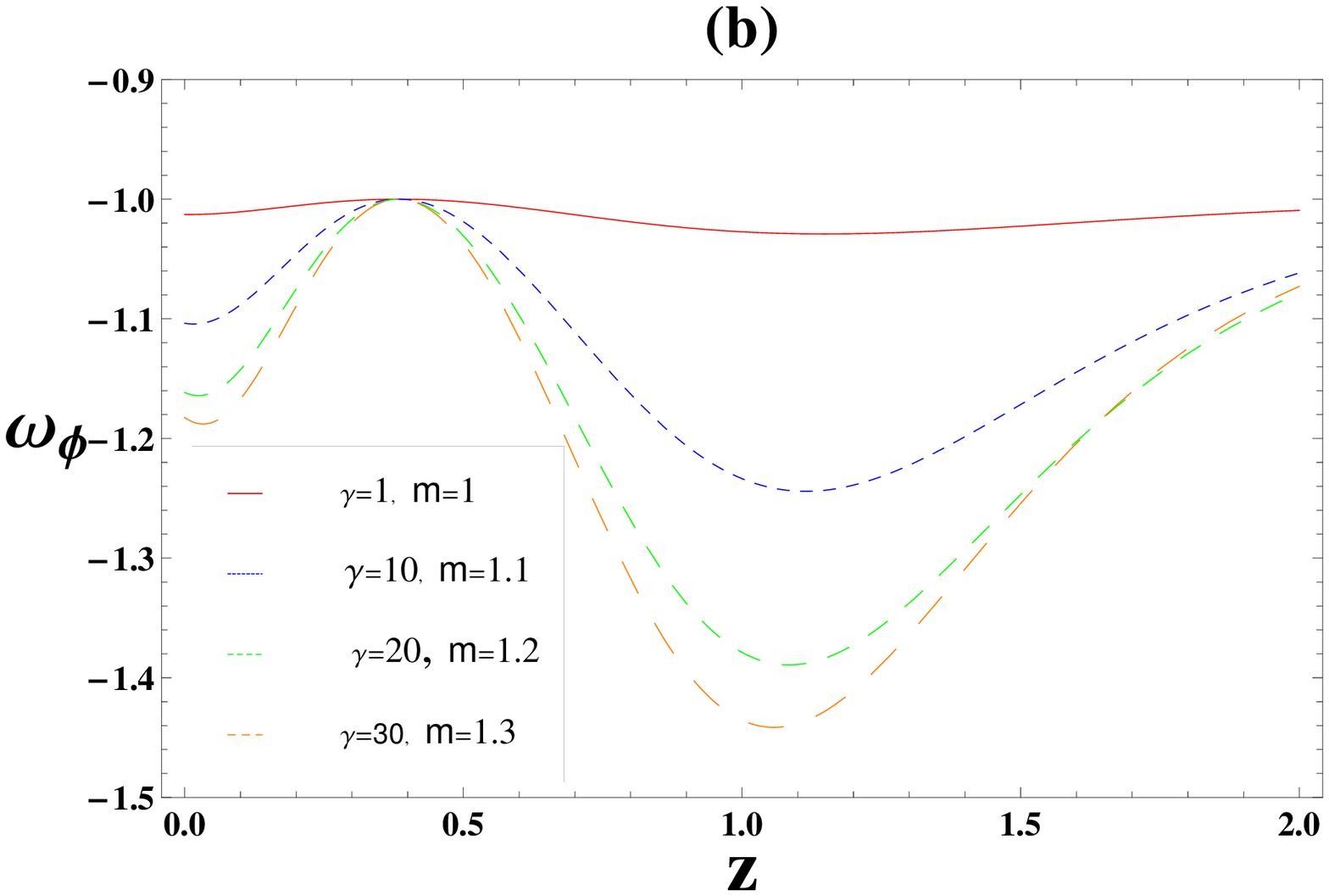}
\caption{The plot of $\omega_{\phi}(z)$ for (a) Starobinsky's and
(b) Hu-Sawicki's models. As the figures indicate, there is a
phantom-like behavior in the latter.}
\end{center}
\end{figure}

\begin{figure}[ht]
\begin{center}
\includegraphics[width=0.43\linewidth]{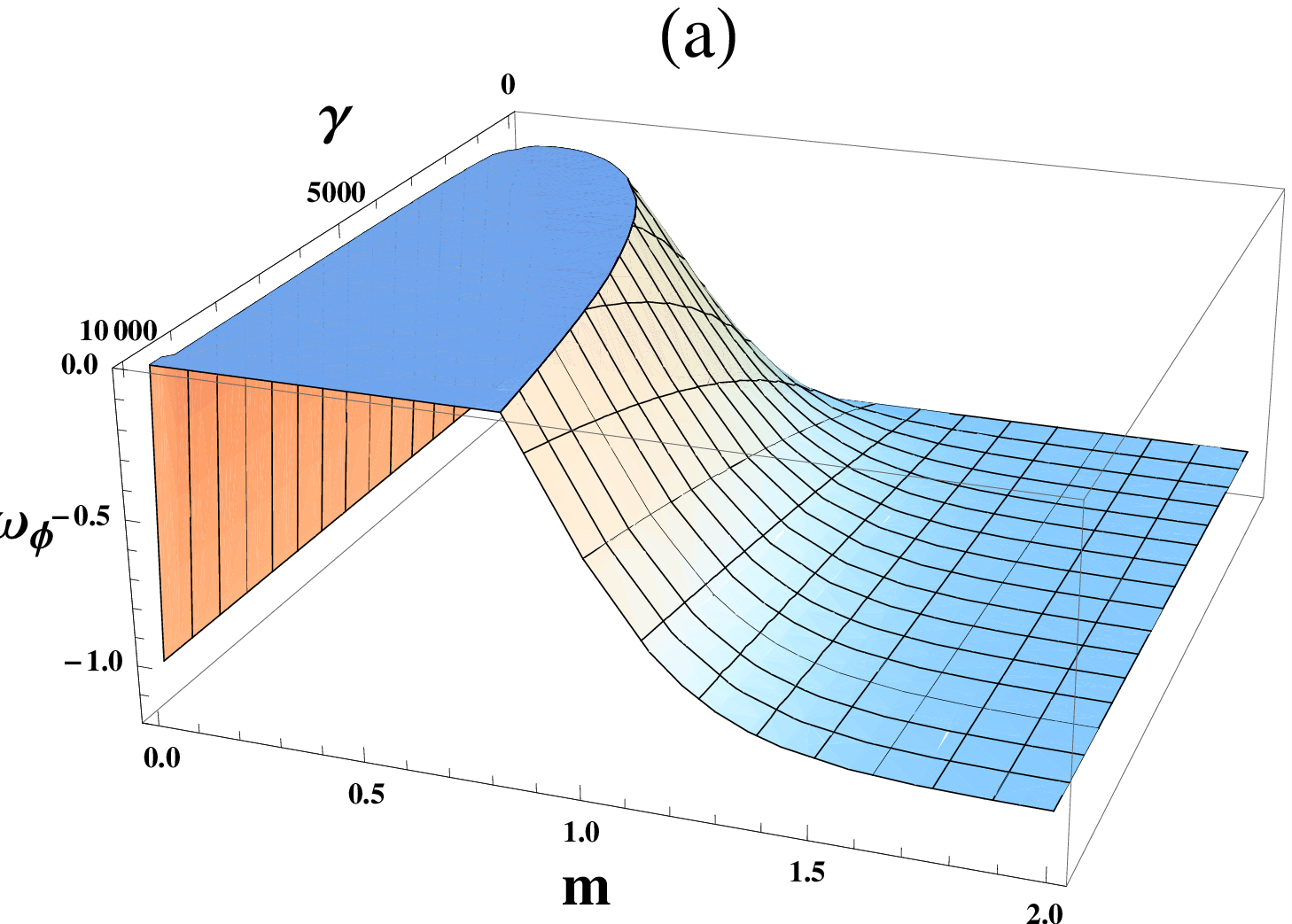}
\includegraphics[width=0.43\linewidth]{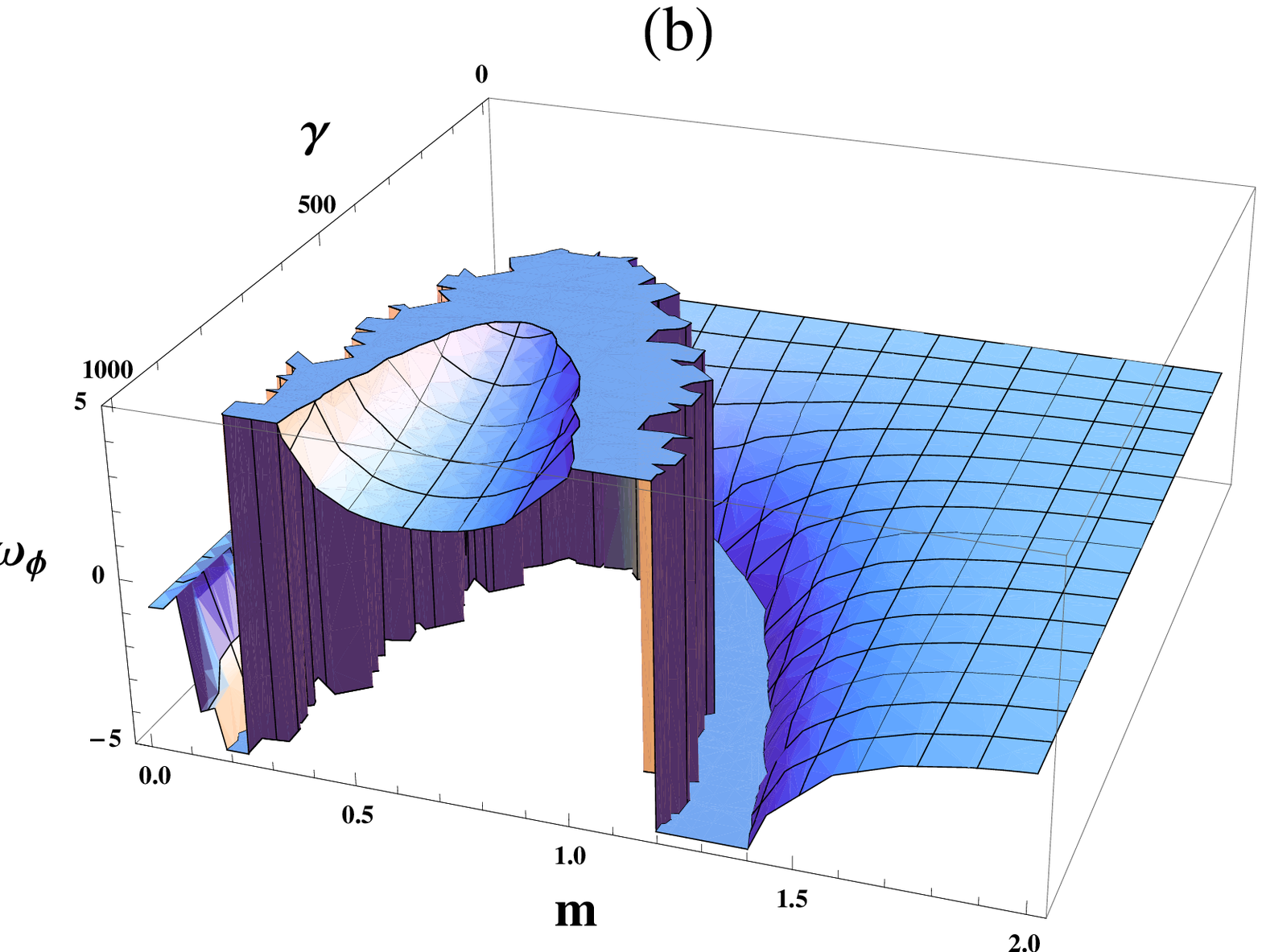}
\caption{The plot of $\omega_{\phi}$ in the redshift $z=0.25$ for
(a) Starobinsky's and (b) Hu-Sawicki's models. As the figures show
the PDL can be crossed in the latter in a small region of the
parameters space.}
\end{center}
\end{figure}

\begin{figure}[ht]
\begin{center}
\includegraphics[width=0.45\linewidth]{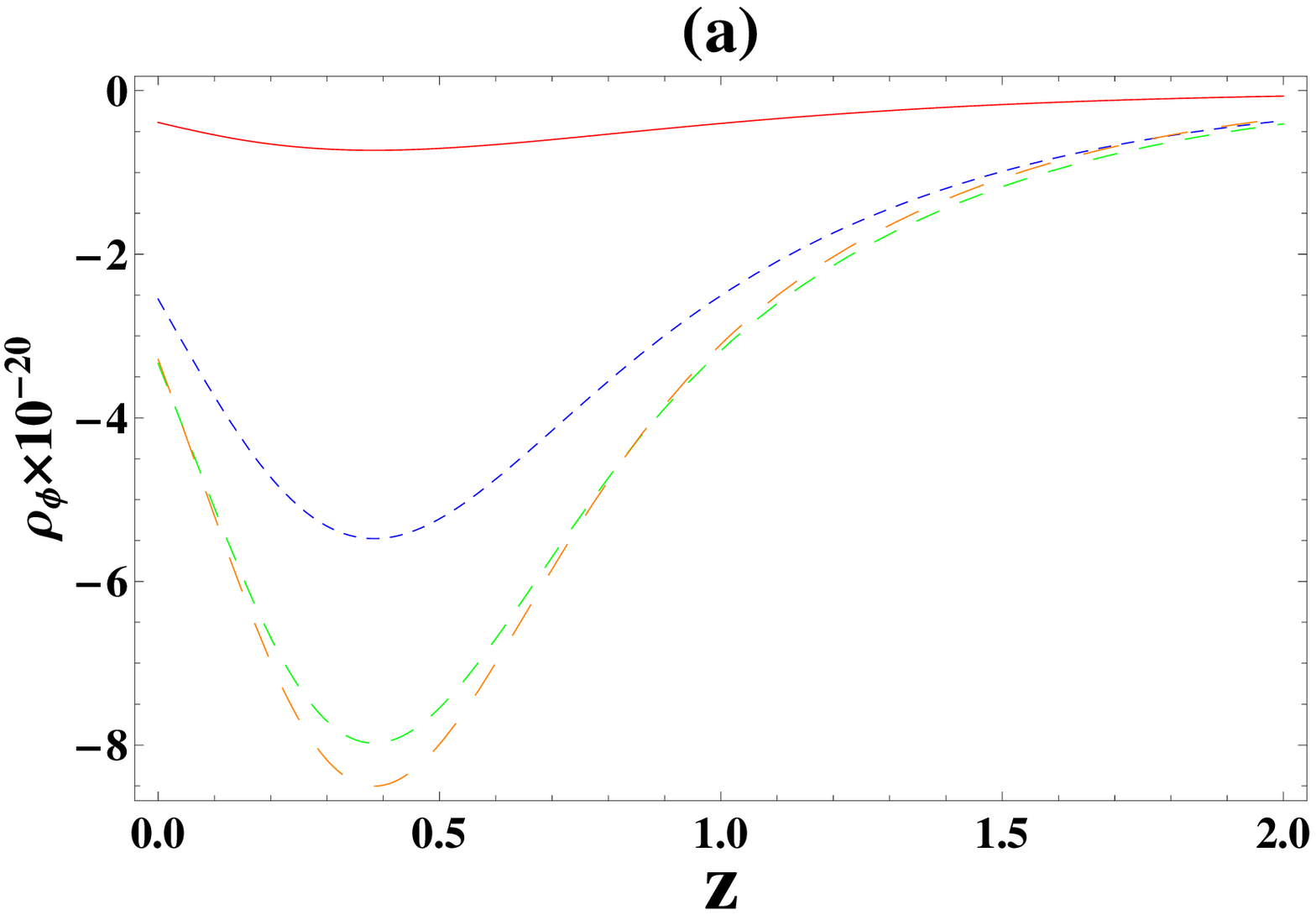}
\includegraphics[width=0.45\linewidth]{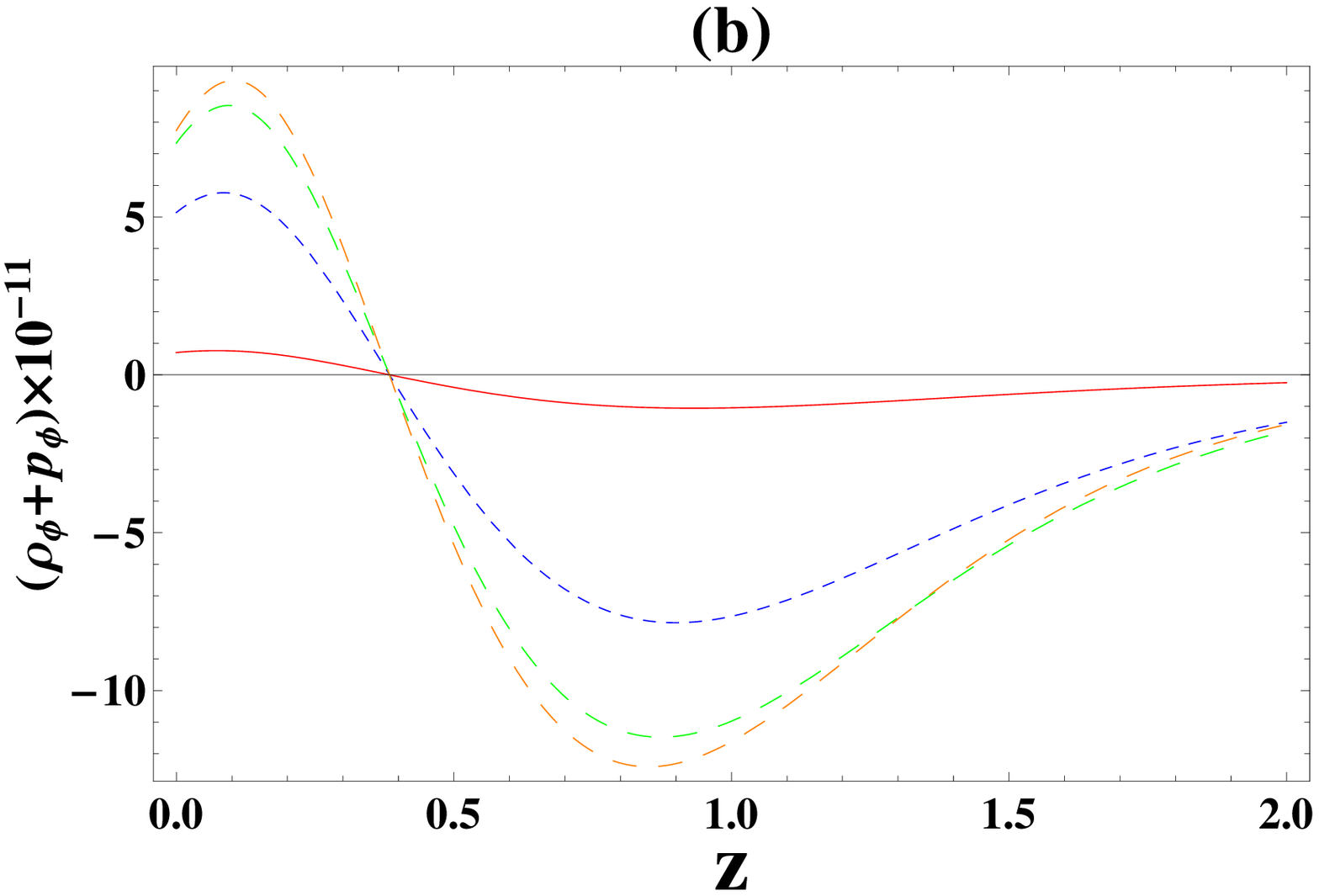}
\end{center}
\end{figure}

\begin{figure}[ht]
\begin{center}
\includegraphics[width=0.45\linewidth]{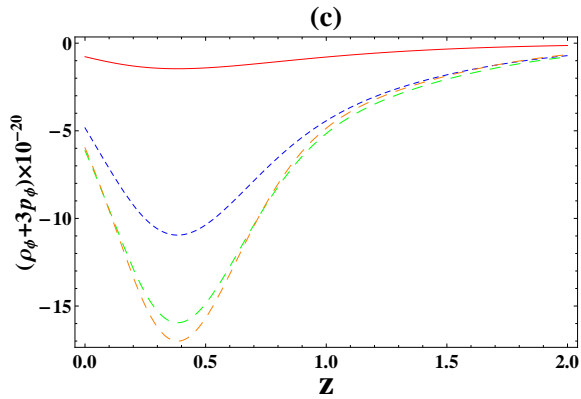}
\caption{Variations of  (a) $\rho_{\phi}$ , (b)
$\rho_{\phi}+p_{\phi}$ and (c) $\rho_{\phi}+3p_{\phi}$ in terms of
the redshift for Hu-Sawicki's model.  The plots indicate that
$\rho_{\phi}< 0$ and $\rho_{\phi}+3p_{\phi}<0$ while
$\rho_{\phi}+p_{\phi}>0$ for $z<0.4$. The curves are plotted for
the same values of the parameters $\gamma$ and $m$ appeared in
Fig.2b.}
\end{center}
\end{figure}

\end{document}